\documentclass[pra,twocolumn,10pt,aps,nofootinbib, superscriptaddress,longbibliography]{revtex4-2}

\usepackage{graphicx}
\usepackage{dcolumn}
\usepackage{bm}
\usepackage{amsmath}
\usepackage{mathtools}
\usepackage{graphicx}
\usepackage{color}
\usepackage{amsthm}
\usepackage{newlfont}
\usepackage{breakcites}
\usepackage{textcomp}
\usepackage{multirow}	
\usepackage{amssymb}
\usepackage{bm}
\usepackage[american]{babel}
\usepackage{braket}
\usepackage{soul}
\usepackage{float}
\usepackage{lipsum}  
\usepackage[normalem]{ulem} 

\usepackage{hyperref}
\hypersetup{
     colorlinks  = true,
     linkcolor    = red,
     citecolor    = cyan,
     urlcolor     = magenta
     }

\begin{document}

\title{Duality of Topological Edge States in a Mechanical Kitaev Chain}

\author{Florian Allein}
\thanks{These authors contributed equally}
\affiliation{Univ. Lille, CNRS, Centrale Lille, Univ. Polytechnique Hauts-de-France, Junia, UMR 8520-IEMN, F-59000 Lille, France}

\author{Rajesh Chaunsali}
\thanks{These authors contributed equally}
\affiliation{Department of Aerospace Engineering, Indian Institute of Science, Bangalore 560012, India}

\author{Adamantios Anastasiadis}
\affiliation{Laboratoire d’Acoustique de l’Universit\'e du Mans (LAUM), UMR 6613, Institut d’Acoustique - Graduate School (IA-GS), CNRS, Le Mans Universit\'e, France}

\author{Ian~Frankel}
\affiliation{Department of Mechanical and Aerospace Engineering, University of California, San Diego, La Jolla, CA 92093, USA}

\author{Nicholas Boechler}
\affiliation{Department of Mechanical and Aerospace Engineering, University of California, San Diego, La Jolla, CA 92093, USA}

\author{Fotios K. Diakonos}
\affiliation{Department of Physics, University of Athens, 15784 Athens, Greece}

\author{Georgios Theocharis}
\email[]{georgios.theocharis@univ-lemans.fr}
\affiliation{Laboratoire d’Acoustique de l’Universit\'e du Mans (LAUM), UMR 6613, Institut d’Acoustique - Graduate School (IA-GS), CNRS, Le Mans Universit\'e, France}

\date{\today}

\begin{abstract}
We theoretically investigate and experimentally demonstrate the existence of topological edge states in a mechanical analog of the Kitaev chain with a non-zero chemical potential. Our system is a one-dimensional  monomer system involving two coupled degrees of freedom, i.e., transverse displacement and rotation of elastic elements. Due to the particle-hole symmetry, a topologically nontrivial bulk leads to the emergence of edge states in a finite chain with fixed boundaries. 
In contrast, a topologically trivial bulk also leads to the emergence of edge states in a finite chain, but with free boundaries. We unravel a duality in our system that predicts the existence of the latter edge states. This duality involves the iso-spectrality of a subspace for finite chains, and as a consequence, a free chain with topologically trivial bulk maps to a fixed chain with a nontrivial bulk.   
Lastly, we provide the conditions under which the system can exhibit perfectly degenerate in-gap modes, akin to Majorana zero modes. These findings suggest that mechanical systems with fine-tuned degrees of freedoms can be fertile test beds for exploring the intricacies of Majorana physics.

\end{abstract}

\maketitle





 
 
 
 




 
 





\textit{Introduction}.---In recent years, topological insulators and superconductors have emerged as candidates for robust and defect-immune manipulation of fermions~\cite{Hasan2010, Qi2011, Bernevig2013, Sato2017}. 
The underlying notion of the topology of band structures, which is intrinsically linked to the concept of geometric phases~\cite{Pancharatnam1956, Berry1984}, has also been used to design novel bosonic systems -- both quantum and classical. These systems have shown robust localization properties at corners, edges, and surfaces~\cite{Cooper2019, Ozawa2019, Susstrunk2016, Ma2019}. 
In classical systems, the exploration of topological physics serves two purposes: (1) Realization of theoretical predictions that are difficult to be observed in electronic or superconducting systems through table-top experimental setups, and (2) The ingredient of topology being utilized for robust manipulation of energy in space, and thus leading to next-generation resilient mechanical logic \cite{Darabi2020} and energy harvesting devices \cite{Chaplain2020}. 
The majority of studies on classical topological analogs has focused on insulators, with significantly fewer exploring superconducting phenomena~\cite{prodan2017dynamical, Guo2021}.

In this work, we take inspiration from the Kitaev chain~\cite{kitaev2001unpaired}, which is an archetypical model of a topological superconductor and has been intensely investigated in quantum systems due to its ability to host topological Majorana zero modes (MZM) \cite{leumer2020exact,jack2021detecting,manna2020signature}. These are expected to be good candidates for quantum computation devices \cite{tutschku2020majorana,aasen2016milestones}. Majorana particles are their own anti-particles, and can manifest in the context of topological superconductors by having exact zero modes that are eigenstates of the particle-hole operator \cite{leumer2020exact}. A Kitaev chain with a non-zero chemical potential undergoes a topological phase transition and topological edge states appear in a specific parameter regime. Within this regime, MZMs appear at discrete 1D lines (or the ``parity switches''~\cite{hegde2016majorana}) in the phase diagram due to finite size effects.

Although mechanical analogues of the Kitaev chain have been investigated in the past
\cite{prodan2017dynamical, Barlas2020, al2021enabling,qian2022observation}, this is the first time that a system with an intrinsic particle-hole symmetry that maps to a non-zero chemical potential Kitaev chain is presented. Furthermore, we find that this system supports a hitherto unobserved
duality between the \textit{finite} lattices with fixed and free boundary conditions.
Not only do the chains with fixed boundaries support edge states for a topologically \textit{nontrivial} bulk as per the bulk-boundary correspondence, but also, the chains with free boundaries support edge states for a topologically \textit{trivial} bulk. We show that this is because a subspace of a fixed chain with $N$ particles is isospectral (dual) to a free chain with $N-1$ particles, under a suitable change of the parameters. The existence and duality of edge states in the system with free boundary conditions is important, as it suggests a hidden, topologically nontrivial origin, which further suggests that they may, in the future, be shown to exhibit topological protection.

Several notions of duality have been investigated in the past~\cite{Kramers1941, Savit1980, Louvet2015, fruchart2020dualities}. However, the duality in this work relates to the isospectral nature of topologically distinct finite chains with different boundaries. This duality does not appear in other mechanical lattices that support finite-frequency edge states, 
such as Su-Schrieffer-Heeger inspired 1D spring-mass model \cite{TheocharisPRB2021}, 
or 1D chains with chiral cells~\cite{kopfler2019topologically, YangArXiv2022}.
We speculate that this duality is connected with the nature of our lattice with coupled degrees of freedom, and the corresponding symmetry group, $\mathbb{Z}_{2}$, which both finite chains support.
Lastly, we investigate theoretically the conditions under which the system can exhibit \textit{perfectly} degenerate in-gap modes (MZMs) in our finite mechanical chain.

\begin{figure}[t!]
\includegraphics[width=\columnwidth]{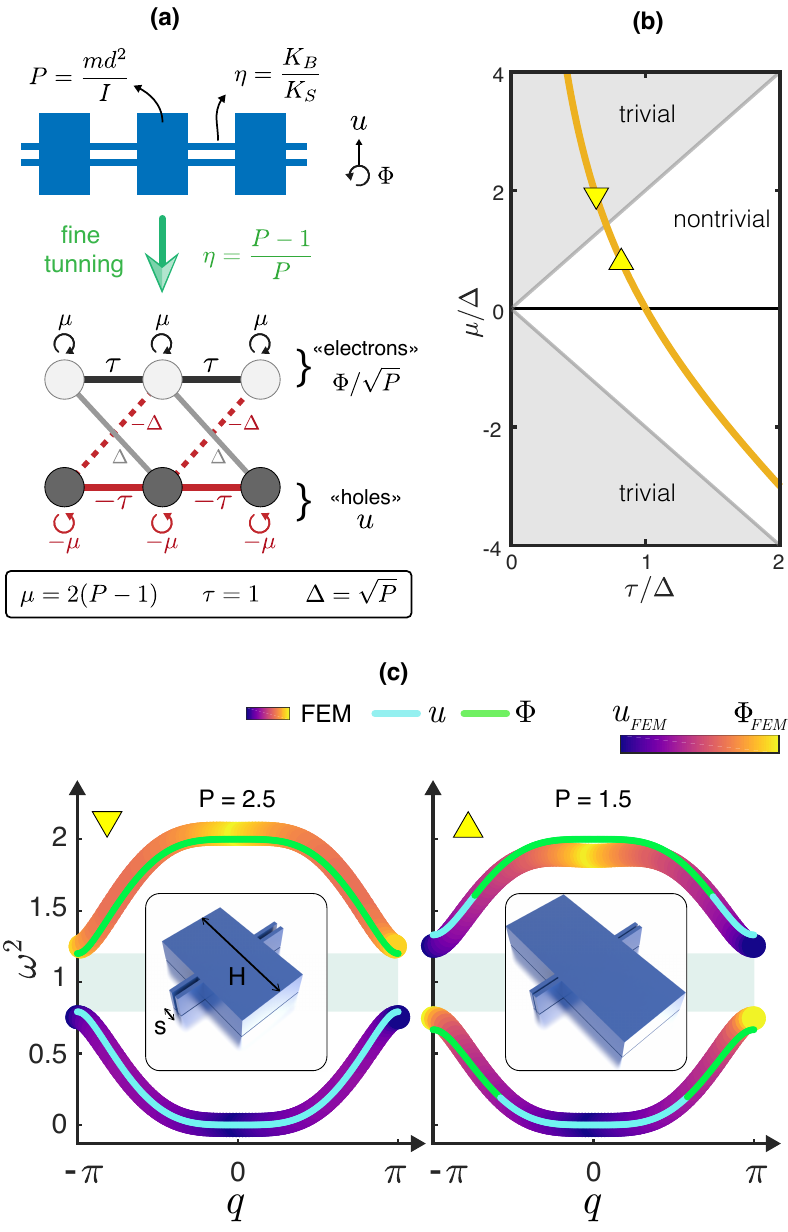}
\caption{\textbf{Mechanical Kitaev chain.} (a) A mechanical monomer chain with transverse and rotational degrees of freedom maps to the Kitaev chain after fine-tuning. 
(b) Topological phase diagram of the Kitaev chain. Our fine-tuned mechanical chain follows the solid yellow line. Two experimental cases $P=1.5$ and $P=2.5$ are marked with triangles.
(c) Dispersion diagrams for $P=1.5$ and $P=2.5$ obtained through both lumped-mass model and finite element method. $H$ and $s$ are the dimensions that are varied. The colorbar denotes the modal dominance.}
\label{fig1} 
\end{figure}

\textit{Kitaev chain}.---A p-wave superconductor widely known as a Kitaev chain \cite{kitaev2001unpaired} is described by the Hamiltonian $H = \frac{1}{2} \Psi^{\dagger}H_{\mathrm{BdG}}\Psi$ in Bogoliubov-de Gennes (BdG) formalism. Here $\Psi$ is a column vector containing all the creation and annihilation operators $\Psi = (c_{1}...c_{N},c_{1}^{\dagger},...c_{N}^{\dagger})$ for $N$ spinless electrons, and

\begin{equation}\label{BdG_k_space}
	H_{\mathrm{BdG}}(q) =
	     \begin{pmatrix}
		-\mu-2\tau \cos{q}  &  -2i\Delta\sin{q}\\
		2i\Delta\sin{q} &  \mu + 2\tau \cos{q}
	  \end{pmatrix},
\end{equation}
where $q\in[-\pi,\pi]$, the lattice constant is unity, $\mu$ is the chemical potential, $\tau$ is the hopping amplitude, and $\Delta$ is the p-wave superconducting pairing constant. By construction, the BdG Hamiltonian anti-commutes with the particle-hole operator $\Pi = \sigma_{1}\mathcal{K}$, where $\sigma_{1}$ is the Pauli matrix and $\mathcal{K}$ accounts for complex conjugation. Furthermore, the BdG Hamiltonian is $\mathcal{PT}$ symmetric, i.e., it commutes with the operator $\sigma_{3}\mathcal{K}$.

\textit{A mechanical analogue of the Kitaev chain}.---In Fig.~\ref{fig1}(a), we show a mechanical structure, whose dynamics is governed by two in-plane degrees of freedom (DOFs): displacement $u$ and rotation $\Phi$. The resulting dynamical matrix has the same form as $H_{\mathrm{BdG}}$ under a specific fine-tuning of parameters. 
In particular, our mechanical system depends on two parameters: The ratio ($P$) of mass to moment of inertia of the lumped masses and the ratio ($\eta$) of bending stiffness to shear stiffness.
To achieve the desired fine-tuning, we impose $\eta = 1 - (1/P)$ (see Supplemental Material \cite{SI} for details). 
Consequently, our system maps to the parameters of the Kitaev chain in the following way: $\Delta \rightarrow \sqrt{P}, \tau \rightarrow 1, \mu \rightarrow 2(P-1)$. 
Since we vary $P$ in our design, we trace a 1D path (with a nonzero chemical potential, in general) in the phase space of the Kitaev chain, as is shown in Fig.~\ref{fig1}(b), and transition between topologically trivial and nontrivial regimes. 

An interesting feature of this mechanical system is that it is a \textit{monomer}, but with two coupled DOFs. If a particle is removed from the chain, two DOFs for that particle will be lost, and in this sense, these DOFs are inseparable. This property resembles the particle-hole property of the Kitaev chain, i.e., a particle cannot be removed from the system without removing a hole at the same time. 
 
\textit{Dispersion analysis}.---In Fig.~\ref{fig1}(c), we show the dispersion curves obtained for different values of parameter $P$ [corresponding to the triangles in Fig.~\ref{fig1}(b)]. These are obtained by calculating the eigenvalues of the dynamical matrix for a unit cell with Bloch-Floquet periodic boundary conditions. In addition, we calculate the modal dominance using the eigenvectors.
We observe two branches in the dispersion diagram as a result of the lumped-mass model having two DOFs, i.e., $u$ and $\Phi$, per mass [we validate our lumped-mass model by performing finite element method (FEM) simulations on the 3D design of the unit cells, which shows an excellent agreement]. We also observe that the entire spectrum ($\omega^2$) is symmetric about a mid axis, which is $\omega^2=1$, after normalizing the dynamical matrix. This is due to the particle-hole symmetry of the dynamical matrix $\tilde{D}$ around a finite frequency $\omega^2=1$, such that 
$\sigma_1\left( \tilde{D} - \bm{I} \right)\sigma_1^T  = - \left( \tilde{D} - \bm{I} \right)$.

Recall that Fig.~\ref{fig1}(b) had suggested a topological transition when $P$ is varied. This is evident when we notice a band gap between the two dispersion branches for $P>2$. As we decrease $P$, the band gap closes at $P=2$ and opens again for $P<2$ (see Supplemental Material \cite{SI} for details).  
Moreover, we observe a clear difference in terms of relative dominance of  $u$ and $\Phi$ for $P>2$ and $P<2$. At the end of the Brillouin zone, i.e., at $q= \pi$, we witness that the dominant motion in dispersion branches is inverted -- indicating a band inversion. 

This band inversion is accompanied by a topological transition in the system. Since the existence of particle-hole symmetry allows us to define a winding number, we can characterize the topology of the bulk. 
We obtain zero winding number for $P>2$ (topologically trivial), and a unit winding number for $P<2$ (topologically nontrivial).

\textit{The bulk-boundary correspondence}.---As per the bulk-boundary correspondence, we expect to witness topologically protected edge modes in the finite lattice made of topologically-nontrivial unit cells. This is true as long as we keep the particle-hole symmetry of the entire finite chain intact. For the mechanical system as ours, the symmetry is obeyed when the chain has fixed boundary conditions on both sides (see Supplemental Material \cite{SI} for the discussion on finite chain). 

\begin{figure}[t!]
\begin{center}
\includegraphics[width=1\columnwidth]{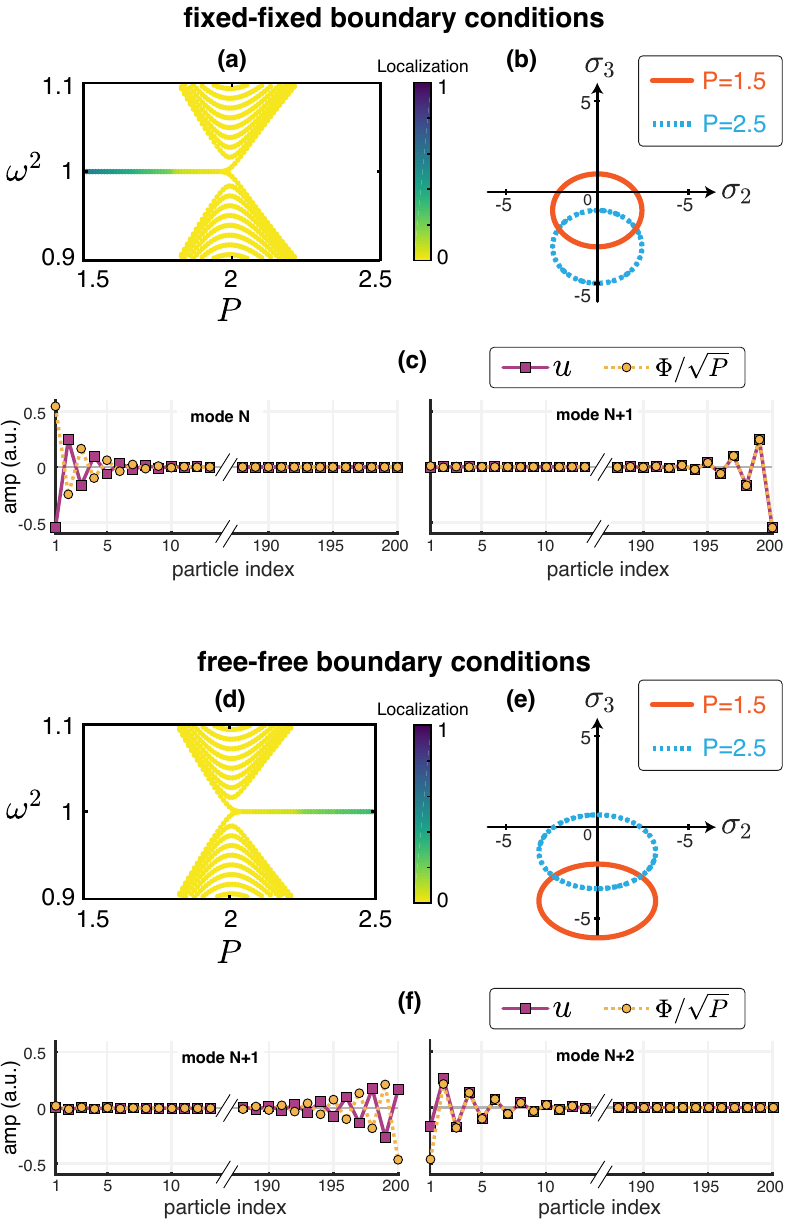}
\end{center}
\caption{\label{fig2} \textbf{Finite chain and edge state duality.} 
(a) Spectrum of a finite chain ($N=200$) with both boundaries fixed as a function of $P$. The colorbar indicates the localization of an eigenstate. (b) Winding in Bloch plane for topologically trivial ($P=2.5$) and nontrivial ($P=1.5$) configurations. (c) Eigenshapes of the localized states on the fixed edges. (d)---(f) The same for a chain with free boundaries. An edge state appears for a topologically trivial bulk ($P=2.5$), highlighting the inherent duality in our system.}
\end{figure}

In Fig.~\ref{fig2}(a), we show the spectrum of a long chain of 200 particles with various values of $P$ for fixed-fixed boundaries. Clearly, we notice the emergence of two localized states inside the band gap for all topologically nontrivial bulk ($P<2$). This is a result of nontrivial winding of the Bloch vector in the $\sigma_2$ - $\sigma_3$~plane, as is shown in Fig.~\ref{fig2}(b). In Fig.~\ref{fig2}(c), we plot these two eigenstates, which are localized on the left and the right end of the chain. 
These topologically-protected edge states have mixed polarization (in terms of displacement and rotation), which is different from the ones seen in the SSH model~\cite{TheocharisPRB2021}.

One may now ask: \textit{What happens to the bulk-boundary correspondence in the case of ``free-free'' boundaries?} Since such a finite chain does not preserve particle-hole symmetry (see Supplemental Material \cite{SI} for the dynamical matrix $D_{bc}$ for a finite-chain), it is not clear how the bulk topology can be related to the existence of edge states. In Fig.~\ref{fig2}(d), we show the spectrum of the chain with varying $P$. Remarkably, we witness the emergence of two edge states inside the band gap in this case as well. However, the edge states exist for $P>2$, i.e., topologically-\textit{trivial} bulk.  
Moreover, the spectrum \textit{appears} to be symmetric around $\omega^2=1$ again. This leads us to investigate the topological origin of these edge states next.

\textit{Duality of topological edge states}.---The spectrum of a ``free-free'' chain with $N$ particles have $2N$ eigenfrequencies. Two of these eigenfrequencies are zero due to free boundary conditions. We show that the subspace of remaining $2N-2$ eigenfrequencies is isospectral to the eigenspace of a chain of $N-1$ particles, but with \textit{fixed} boundary conditions and the transformation: $P_{\text{free}} \rightarrow P_{\text{fixed}}/(P_{\text{fixed}}-1)$ (see Supplemental Material \cite{SI} for a test case and the proof). Therefore, the spectrum of a \textit{free-free} chain is completely determined by examining the spectrum of a \textit{fixed-fixed} chain with less number of particles and modified $P$. 
In this way, the emergence of edge states of the free-free chain with $P_{\text{free}}>2$ can be explained from the bulk-boundary correspondence of the fixed-fixed chain with $P_{\text{fixed}}<2$ since $P_{\text{free}} \rightarrow P_{\text{fixed}}/(P_{\text{fixed}}-1)$. This highlights the duality of topological edge states in our system. In Fig.~\ref{fig2}(e), we show that $P>2$ for the free chain maps to $P<2$ for the fixed chain, which have a nontrivial winding. In Fig.~\ref{fig2}(f), we plot the edge states emerging at the free boundaries. Their shapes are different compared to the ones in the fixed-fixed case, in that the $u$ and $\Phi$ at each site are not \textit{exactly} the same or opposite. Also, ($N+1$)th and ($N+2$)th modes are right and left localized, respectively, as compared to the fixed-fixed chain, where $N$th and ($N+1$)th modes are localized on the left and right ends, respectively. Such a difference in eigenstates persists even for the free-free chain and its \textit{dual} fixed-fixed chain, and is connected to the fact that the symmetry groups of the $N$-particle mechanical chain with fixed and free ends, are different
(see Supplemental Material \cite{SI} for details).

\begin{figure}[t]
\begin{center}
\includegraphics[width=\columnwidth]{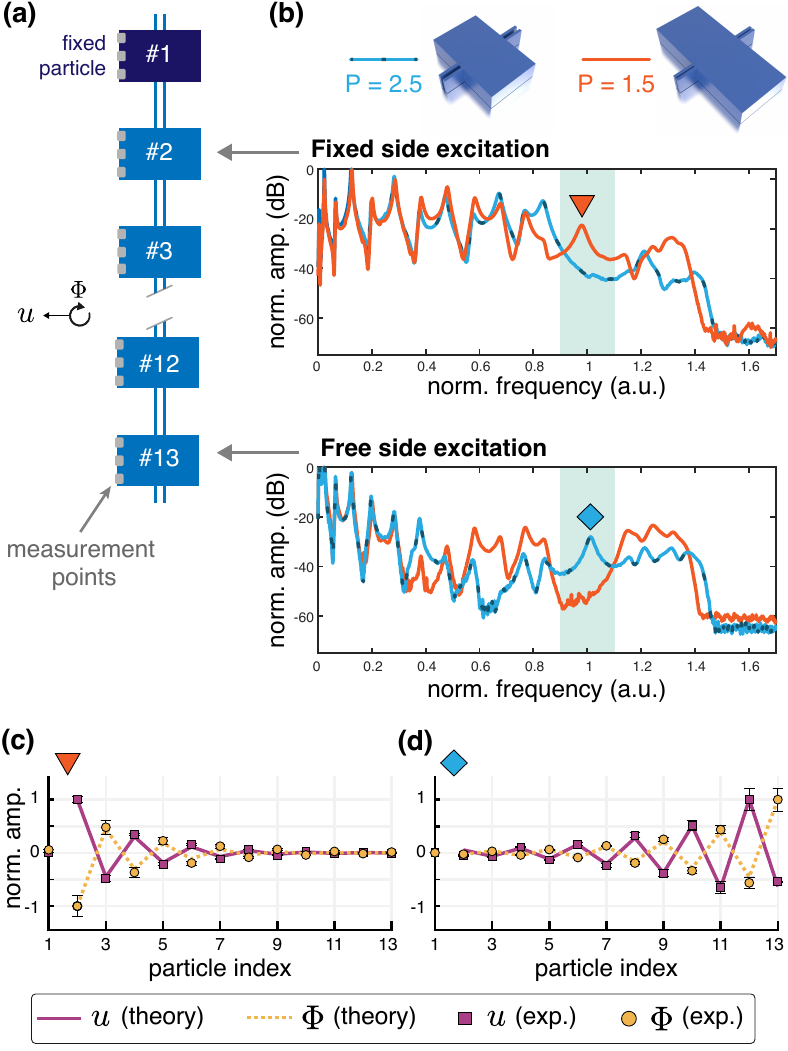}
\end{center}
\caption{\label{fig3} 
\textbf{Experimental observation of edge states.}
(a) Schematic of the experimental setup, which is suspended vertically by fixing particle $\sharp$1.
Three points are probed in each particle to characterize the transverse displacement and rotation.
(b) Experimental frequency response at particle $\sharp$7 when the chains with $P=1.5$ and $P=2.5$ are excited at the fixed end (at particle $\sharp$2) or at the free end (particle $\sharp$13). The blue area corresponds to the band gap. 
Edge state (c) at the fixed boundary for $P=1.5$, and (d) at the free boundary for $P=2.5$.}
\end{figure}

\textit{Experimental observation of edge states}.---To experimentally verify the existence of edge states in both fixed and free ends of our mechanical chain, we prepare our test setup with fixed-free boundary conditions so that both types of edge states can be observed in the system without changing the mounting. For a large chain with a negligible interaction between two boundaries, we expect the emergence of an edge state at the fixed end is governed by the bulk-boundary correspondence of fixed-fixed chain. Similarly, we expect an edge state at the free end as well, albeit for different $P$ values than the fixed chain, governed by the duality in free-free chain mentioned earlier. This follows that the fixed-free chain should always have an edge state for all values of $P$ except $P=2$, for which the band gap closes. For $P<2$, it would support an edge state on the fixed end, whereas for $P>2$, it supports an edge state on the free end.

We fabricate chains of 13 masses (large cuboids) through additive manufacturing and suspend them vertically by mounting the particle $\sharp1$ as shown in Fig.~\ref{fig3}(a). Therefore, the system represents a fixed-free chain. We consider two chains with different $P$ and excite them with an automatic modal hammer by striking the particle $\sharp2$ or $\sharp13$ corresponding to the fixed or free sides. By using a laser Doppler vibrometer, we then measure the velocity at multiple points along the chain. See Supplemental Material~\cite{SI} for more details on fabrication, experimental setup, and data acquisition. 

Figure~\ref{fig3}(b) shows experimentally obtained frequency response at particle $\sharp7$ when the chains with $P=2.5$ and $P=1.5$ are excited from different ends. We witness a band gap (highlighted region) and a peak inside it in some cases. In particular, the state inside the band gap exists at the fixed end for $P=1.5$, and at the free end for $P=2.5$, as theoretically predicted above.


To verify that these modes are indeed localized at different edges, we reconstruct the mode shapes from the experimental data in Figs.~\ref{fig3}(c),(d). We observe an excellent agreement between the theory and experiments in predicting amplitude decay as one goes away from the boundaries. 
We also note that the edge state localized at the free end [Fig.~\ref{fig3}(d)] is different in its shape compared to its counterpart for the fixed edge, as discussed earlier. Therefore, the observation the edge states for both fixed and free ends demonstrates the presence of underlying duality in our mechanical system.

\begin{figure}[!]
\begin{center}
\includegraphics[width=1\columnwidth]{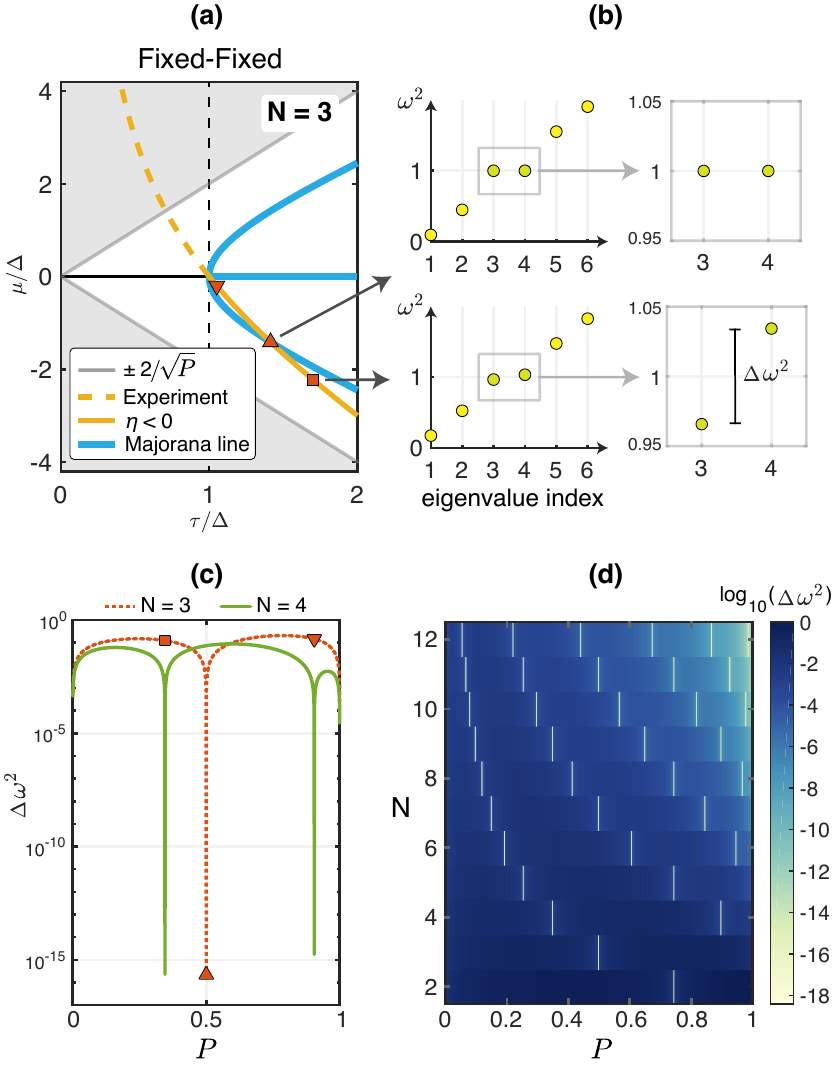}
\end{center}
\caption{\label{fig4} \textbf{Majorana zero modes in a fixed-fixed chain.} 
(a) Kitaev phase diagram and Majorana lines for a short chain with $N=3$. Our fine-tuned model can lie on the Majorana lines for $P<1$ and $\eta<0$. 
(b) The mid gap states are perfectly degenerate when the system falls on the Majorana lines (triangle) as opposed to the case away from the Majorana lines (square).
The perfect degeneracy in mid-gap states in (c) short chains of length $N=3$ and $N=4$, and (d) chains with various values of $N$.}
\end{figure}

\textit{Majorana zero modes (MZMs)}.---An interesting feature of the Kitaev chain is that it exhibits Majorana zero modes. In Fig.~\ref{fig4}(a), we show the MZMs are represented as 1D curves (thick lines) in the phase diagram of the Kitaev chain (see Supplemental Material \cite{SI} for the calculations). 
If the system lies exactly on these lines, the edge states exhibit a perfect degeneracy ($\Delta \omega^2 =0$) even for a finite chain ($N=3$) as shown in Fig.~\ref{fig4}(b). Our fine-tuned mechanical system lies on the yellow line of the Kitaev phase diagram shown in Fig.~\ref{fig4}(a). In order to be exactly on the Majorana lines, we must have negative $\mu$. This translates to having $P<1$ and $\eta<0$. 
In Figs.~\ref{fig4}(c),(d), we include numerically-obtained spectra for fixed-fixed chains of various lengths and $P<1$. We witness sharp dips in $\Delta \omega^2$, therefore highlighting the possibility of perfect degeneracy of edge states in short mechanical chains.

For the experimental realization of such scenario, we would need either negative bending or negative shear stiffness in our mechanical chain. 
We suggest that a dynamic negative stiffness may be implemented through the use of local resonance~\cite{Huang2011, Matlack2018}. If the gap created by the bending and shear stiffnesses is sufficiently large, then, providing sufficiently low damping, an attached local resonator may be able to induce an effective negative stiffness in the frequency vicinity of the edge modes, but minimally affect the band structure defined in the absence of local resonance. 
Interestingly, duality of edge states in our system suggests that such MZMs are also possible at the free ends of the chain. By following the required mapping, we deduce that a free chain would require a negative effective \textit{mass} to have perfectly degenerate edge states. This may also be achieved experimentally by the use of local resonators.

\textit{Acknowledgments}.---We thank N. Herard, B. Skoropys, and M. Coimbra for earlier investigation on the experimental realization of a mass-spring system via additive manufacturing.
R.C. acknowledges the Startup Grant provided by the Indian Institute of Science.
I.F. acknowledges support from the NDSEG Fellowship Program through the US Army Research Office.
N.B. acknowledges supported from the US Army Research Office (Grant No. W911NF-20-2-0182).

\def\bibsection{\section*{}} 

\bibliography{bibliography.bib}

\end{document}